\documentclass[11pt,preprint]{aastex}

\begin{document}

\def\lapp{\ifmmode\stackrel{<}{_{\sim}}\else$\stackrel{<}{_{\sim}}$\fi}
\def\gapp{\ifmmode\stackrel{>}{_{\sim}}\else$\stackrel{>}{_{\sim}}$\fi}
\def\kms{km~s$^{-1}$}
\def\psr{PSR~B1757$-$24}
\def\snr{G5.4$-$1.2}

\title{X-ray Detection of Pulsar PSR B1757$-$24 and its
Nebular Tail}

\author{
V. M. Kaspi,\altaffilmark{1,2,3}
E. V. Gotthelf,\altaffilmark{4}
B. M. Gaensler,\altaffilmark{2,5}
M. Lyutikov\altaffilmark{1,6}
}

\altaffiltext{1}{Department of Physics, Rutherford Physics Building,
McGill University, 3600 University Street, Montreal, Quebec,
H3A 2T8, Canada}

\altaffiltext{2}{Department of Physics and Center for Space Research,
Massachusetts Institute of Technology, Cambridge, MA 02139}

\altaffiltext{3}{Canada Research Chair, Alfred P. Sloan Fellow}

\altaffiltext{4}{Columbia University Astronomy Department, Pupin Hall,
550 West 120th Street, New York, NY 10027}

\altaffiltext{5}{Hubble Fellow; current address Harvard-Smithsonian
Center for Astrophysics, 60 Garden Street, Cambridge, MA 02138}

\altaffiltext{6}{CITA National Fellow}



\begin{abstract}

We report the first X-ray detection of the radio pulsar PSR B1757$-$24
using the {\it Chandra X-ray Observatory}.  We detect
point-source emission at the pulsar position plus a
faint tail extending nearly 20$''$ east of the pulsar, in the same
direction and with comparable morphology to the radio tail.  Assuming
the point-source X-ray emission is magnetospheric, the observed X-ray
tail represents only $\sim$0.01\% of the pulsar's spin-down
luminosity.  This is significantly lower than the analogous
efficiencies of most known X-ray nebulae surrounding rotation-powered
pulsars. Assuming a
non-thermal spectrum for the tail photons, we show that the tail is unlikely to be emission left behind following the
passage of the pulsar, but rather is probably from synchrotron-emitting pulsar
wind particles having flow velocity $\sim$7000~km~s$^{-1}$.  We
also show that there must be a significant break in the tail
synchrotron spectrum between the radio and X-ray bands that is
intrinsic to the particle spectrum.  No emission is detected from the
shell supernova remnant G5.4$-$1.2.  The upper limits on remnant
emission are unconstraining.

\end{abstract}


\section{Introduction}
\label{sec:intro}

\psr\ is a 124-ms radio pulsar discovered near the supernova
remnant \snr\ \citep{mdt85}.  Radio timing observations established
that the pulsar has characteristic age 16~kyr and spin-down
luminosity $\dot{E} = 2.6 \times 10^{36}$~erg~s$^{-1}$ \citep{mkj+91}.  The
pulsar is at the tip of a flat-spectrum radio protuberance 
just outside the west side of the supernova remnant (SNR) shell.
The protuberance consists of a small, roughly circular nebula,
G5.27$-$0.90, from which a collimated jet-like feature is emerging on its western side.
The pulsar is at the westernmost tip of the collimated feature.
The morphology and spectrum
of the jet are strongly suggestive of a ram-pressure
confined pulsar wind nebula (PWN) \citep{fk91}.  
Thus, assuming an association between the pulsar and \snr, the former
appears to have overtaken the expanding shell, presumably because
of a large kick at birth.  Given the angular 
displacement of the pulsar from the best-estimate remnant center, and
if the pulsar's characteristic age is a good estimate of its true
age, this implies a proper motion of $\sim$75~mas~yr$^{-1}$, corresponding
to a transverse space velocity of $v_t \sim$1800~km~s$^{-1}$ \citep{fkw94}, for a
distance of 5~kpc \citep{ckk+87,fkw94}.  
However, recent interferometric observations have failed to
detect the implied proper motion \citep{gf00}.  They set a 5$\sigma$
upper limit on the proper motion of $< 25$~mas~yr$^{-1}$,
corresponding to $v_t < 590$~km~s$^{-1}$.  This suggests
that the pulsar is older than its characteristic age, or that
the assumed pulsar birth place is incorrect.

A ram-pressure confined pulsar wind should radiate
X-rays as part of the broad-band synchrotron spectrum that
results from the shock-acceleration and subsequent
gyration of relativistic wind electron/positron pairs in the
ambient magnetic field.  
We report here on {\it Chandra X-ray Observatory} observations of
\psr\ in which we detect the source for the first time in X-rays.

\section{Observations and Results}
\label{sec:obs}

The \psr/\snr\ field was observed using the Advanced CCD Imaging
Spectrometer (ACIS) instrument aboard the {\it Chandra X-ray
Observatory} on 2000 April 12.
The ACIS CCD camera \citep{bgb+97} is sensitive to X-rays in the
0.2--10~keV band with 0\farcs492 pixels.
The pulsar was positioned on the
back-illuminated S3 chip of the ACIS-S array, offset from the aimpoint
by $-$0\farcm37 in the Y-direction, in order to avoid source counts from
being split between two readout nodes. Data were collected
in the nominal timing mode, with 3.241~s exposures between CCD
readouts, and ``FAINT'' spectral mode. The standard {\it Chandra} screening
criteria produced a total usable exposure time of 19.6~ks.

\begin{figure}
\plotone{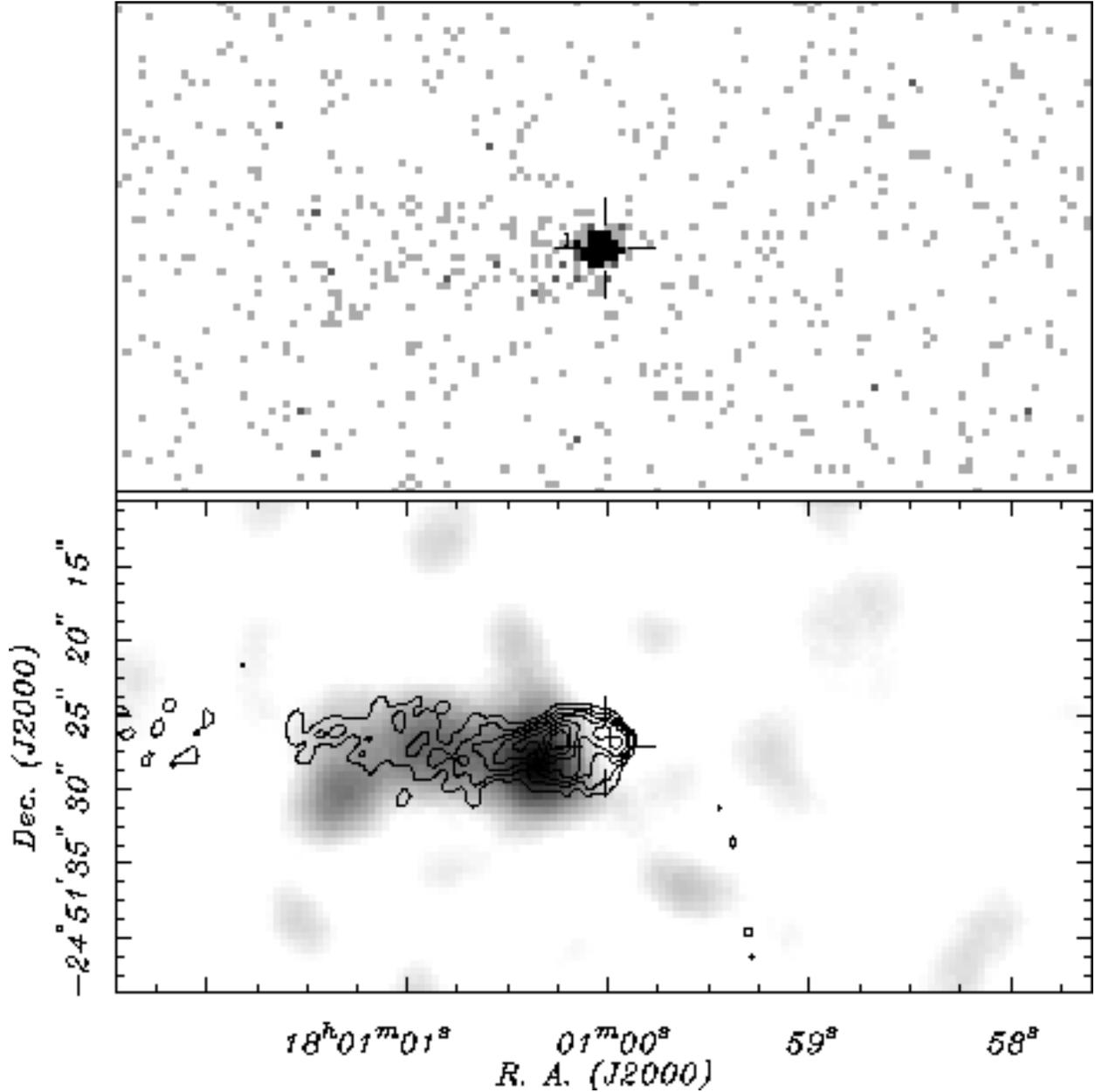}
\figcaption{{\it Chandra} X-ray detection of PSR~B1757$-$24 and
its nebular tail.
Top panel:  greyscale plot of the $0.3-10$~keV
image, centered on the pulsar, and scaled to show all pixels with
detected photons. The X-ray point source is coincident with
the radio pulsar position, denoted by the cross. 
The contours show 4.9~GHz emission in equally spaced contours
ranging
from 0.17 to 0.56~mJy/beam.  The
radio data have beam size 1$''$.2 and
were obtained at the VLA on 1999 October 23 by one
of us (B. Gaensler).
Bottom panel:
same image with the X-ray point source contribution removed, the
image smoothed with a Gaussian of width $\sim 2''$, and the intensity
scaled to emphasize the diffuse X-ray emission. \label{fig:image}}
\end{figure}

\begin{figure}
\plotone{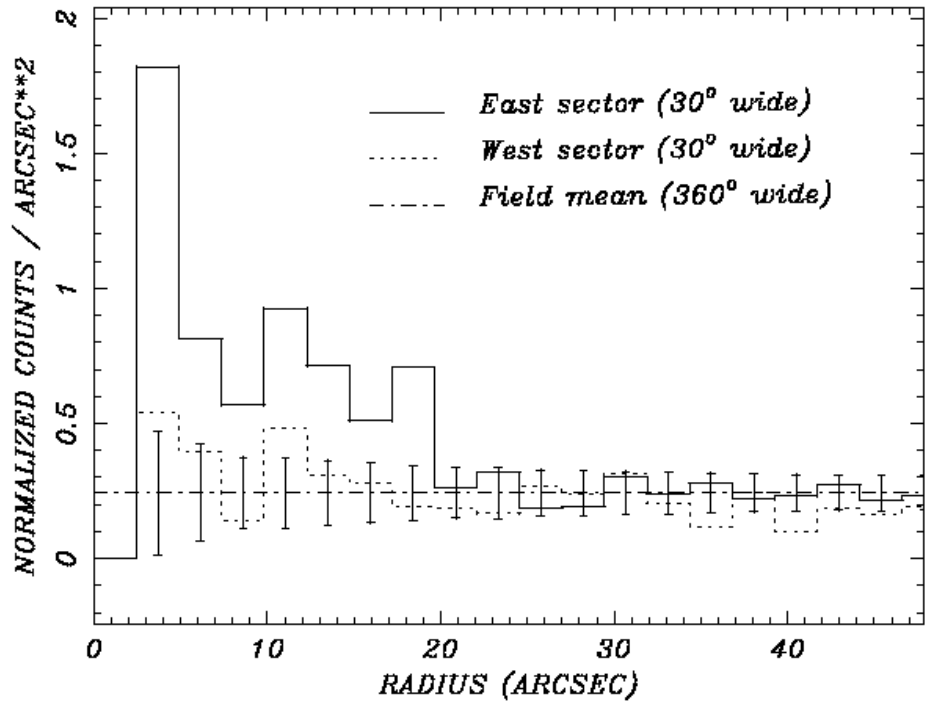}
\figcaption{X-ray intensity as a function of radius from the point source
in two $30^{\circ}$ wedges, one east (solid line) and one west (dotted line)
of the pulsar.  The
point-source emission has been subtracted from the plot.  The field
mean is shown by a dot-dashed line.  $1\sigma$ error bars for each
radial bin are shown.
\label{fig:psf}}
\end{figure}

The ACIS image reveals a faint point source near the radio pulsar position,
and a fainter tail of emission on the eastern side, similar to the
radio emission (Figure~\ref{fig:image}).  Figure~\ref{fig:psf} shows the
distribution of counts as a function of distance from the point source
(with the point source excluded) in two 30$^{\circ}$ wedges on the
eastern and western sides of the pulsar, along with the mean of the
field.  With respect to the mean and the western side, the eastern side
clearly has an overdensity of counts extending nearly 20$''$ from the
point source.  Simulations of a point source at the pulsar position
using the {\it Chandra} simulator {\tt MARX}
are consistent with what is observed in all directions but east.  As
the tail is in the chip readout direction, we were
concerned about it being an instrumental artifact.  However, trails in
ACIS images aligned with the readout direction are only seen for bright
sources, are symmetric about them, and have uniform intensity along
the readout axis (M. Bautz, D. Edgar, H. Marshall, personal
communication).  Furthermore, we verified that there were no
pointing anomalies during the observation and that the count rate
for the tail photons is uniform in time over the observation. We
therefore conclude that the extended, collimated emission we have detected is genuinely
astrophysical.

The standard processing for ACIS produces measured source coordinates
with a nominal accuracy quoted as 0\farcs6, however several problems are
known to exist with the plate scale solution at the time of
writing\footnote{http://asc.harvard.edu/mta/ASPECT/aspect\_caveats.html}.
We derived our own astrometric solution by comparing field stars in
our observation with catalogued optical coordinates. The X-ray source
positions were measured using the CIAO source detection software {\tt
celldetect} with the cleaned $0.3-10$ keV Level 2 event file as
input. For a signal-to-noise threshold set to 3 and the background
estimated using {\tt ldetect}, we found 9 significant X-ray
sources within 5$'$ of the pulsar position.  The on-line
USNO-A2.0 astrometric reference catalogs \citep{mbc+96b}
provided likely counterparts for 8 of the 9 X-ray sources.
The USNO catalog gives stellar positions
to an nominal $1\sigma$ error of $\approx 0\farcs25$ \citep[][Monet, personal communication]{deu99} 
although proper motion has not been applied.  
The offsets between the
optical and X-ray positions are well clustered,
strongly suggesting
correct identifications.  The average offset is
$\sim 0\farcs$6, consistent with the expected systematic
error in the absolute coordinates of ACIS.
The calibrated position of the putative X-ray counterpart to the
pulsar is RA $18^{\rm h}
01^{\rm m} 00^{\rm s}.027$, DEC $-24^{\circ} 51^{\prime} 27\farcs53$
(J2000) with RMS error $0\farcs31$. This is within
1.2$\sigma$ of the radio pulsar location found by \citet{gf00}.  Thus, the point source is
consistent with being the X-ray counterpart of the pulsar.

To characterize the energy dependence of the emission from the point
source, a spectrum was extracted from a circular region having radius
7 pixels (3\farcs4).  An annular background region was chosen
(approximately centered on the point source) having inner and outer
radii 33 and 98 pixels, respectively.  We found a total of 438 source counts and
an estimated 8 background counts in the source region.  
We note that the background on the S3 chip was dominated by scattered emission from
the nearby bright X-ray binary GX~5$-$1.
The CIAO tools
{\tt mkrmf} and {\tt mkarf} were used to generate common response
files for the source and background regions.  The data were regrouped
such that spectral bins contained a minimum of 15 counts, resulting in
27 independent spectral bins.  We fit the spectrum with the
spectral analysis package {\tt
XSPEC} (version 11).
We characterize the spectrum using two different models: a power law
and thermal bremsstrahlung, each photoelectrically absorbed.
Both provide acceptable fits, although the plasma temperature
in the thermal model is poorly constrained.  The results are shown in
Table~\ref{ta:spectrum}.
We verified using {\tt XSPEC}
that the expected $\sim$2.6\% pileup fraction does not have a
significant impact on the fitted parameters.

The tail east of the pulsar is much fainter than the point source.
We attempted to characterize its spectrum by assuming a power-law
model and holding the equivalent neutral hydrogen column density $N_H$
fixed at the value determined from the point source (see
Table~\ref{ta:spectrum}).  We extracted a spectrum for the tail using
a box having dimensions 40$\times$18~pixels, which contained
118~counts.  We used the same annulus to estimate the background as we
did for the point source; in this way we estimate that approximately
40 counts were from the background, leaving 78 source counts.
Spectral fits to the tail in the 0.3--9~keV range for a
power-law model yielded a
mediocre
fit to the data ($\chi^2_{\nu} = 2.0$ for 4 degrees of freedom).  The
best-fit power-law photon index of $1.0 \pm 0.6$ is consistent with
relatively hard emission, as implied by the existence of a significant
number of counts above 5~keV.  However, given the 90\% confidence
range for $N_H$ (Table~\ref{ta:spectrum}), we found that photon
indexes in the range $-$0.32 to 1.8 are consistent with the data.
Lower values of $N_H$, corresponding to lower values of the
photon index, give slightly better fits.  Improved characterization of the
spectrum of the tail must await a deeper observation of the source.
Nevertheless, we can determine the absorved surface brightness in the 2--8~keV band
to be $4.5 \times
10^{-16}$~erg~s$^{-1}$~cm$^{-2}$~arcsec$^{-2}$, with uncertainty
of $\sim$30\%.  The total unabsorbed flux in the 2--8~keV band in our
extraction region is $9.8 \times 10^{-14}$~erg~s$^{-1}$~cm$^{-2}$,
with similar uncertainty.

\section{Discussion}
\label{sec:disc}

\subsection{The Pulsar}

Given the spatial coincidence of the X-ray point source with the radio
pulsar position, the former is likely to be emission
from the radio pulsar itself, in particular, non-thermal pulse-phase-averaged
magnetospheric emission.  
The observed 2--10~keV flux (Table~\ref{ta:spectrum}) implies an
unabsorbed luminosity, for a distance $d=5$~kpc, of $2 \times
10^{33}$~erg~s$^{-1}$, assuming beaming angle $\phi = \pi$~sr.  
This implies an efficiency of conversion of
spin-down luminosity into magnetospheric emission of
$0.00020(\phi/\pi \; {\rm sr})(d/5 \; {\rm kpc})^2$.
This efficiency, as well as the measured power-law photon index
(Table~\ref{ta:spectrum}), are consistent with those 
observed for the magnetospheric components
of other radio pulsars \citep{bt97}.  
The detection of low duty-cycle X-ray pulsations could unambiguously confirm
this interpretation.
The current data cannot however rule out the possibility that the emission is
coming from a very compact nebular region near the pulsar. 
At radio wavelengths, the head of the bow shock is only $\sim 1.5''$ from
the pulsar \citep{gf00}.  The reverse shock, which terminates the wind, is likely
unresolvable in the Chandra image.

\citet{mlr01} suggested that
\psr\ has a fallback disk from which it accretes,
providing non-magnetic dipole spin-down torque to account for the
possible difference between the pulsar's true and characteristic ages.  
They predict a thermal bremsstrahlung spectrum
having $kT \sim 50$~keV and an X-ray luminosity (in an unspecified
energy band) roughly an order of magnitude less than that predicted for
magnetospheric emission.  The {\it Chandra} detection we report here
is consistent with standard magnetospheric emission.
However, our data cannot rule out a fainter thermal component.

\subsection{The Nebular Tail}

The tail X-ray emission is likely to be synchrotron radiation
from the shocked pulsar wind.  
The pulsar wind 
appears to be confined by the ram pressure
of the interstellar medium.  If the pulsar's
space velocity greatly exceeds the ambient medium's sound speed,
a strong forward bow shock travels in front of it, and a reverse
shock closer to the pulsar terminates its relativistic wind.

Conventional wisdom suggests that X-ray synchrotron nebulae should
be smaller than the corresponding radio nebulae because radio
electrons
have much longer lives than do X-rays.  This is consistent with
the apparent contraction of the Crab nebula with increasing photon
energy.  
However, for \psr, Figures~\ref{fig:image} and \ref{fig:psf} demonstrate
that the observed X-ray tail extends nearly 20$''$ (0.48($d$/5~kpc)~pc)
to the east of the pulsar,
nearly as long as the detected radio tail (which, $\sim 10''$ further east,
suddenly
expands to form the flat-spectrum bubble, G5.27$-$0.90;
\citeauthor{fk91}~\citeyear{fk91}).
The $5\sigma$ upper limit on the proper motion implies
a transverse velocity of $v_t <590$~km~s$^{-1}$ \citep{gf00}.  Thus, the time since the 
pulsar was at the eastern-most
tip of the observed X-ray emission must be $> 800$~yr.  The
synchrotron
lifetime of a photon of energy $E$ (in keV) in a magnetic field
$B_{-4}$ 
(in units of $10^{-4}$~G) is $t_s \simeq 40 E^{-1/2}
B_{-4}^{-3/2}$~${\rm yr}$.  Thus, for $t_s > 800$~yr and $E\simeq 1-9$~keV,
$B< 0.8-14 \; \mu$G.  This is much less than the 
equipartition magnetic field  $B_{\rm eq} \sim 
\sqrt{\dot{E}/r_s^2 c } \sim 70 \; \mu$G expected
in the vicinity of the pulsar.  Here, $r_s$ is the distance from the
pulsar to the bow shock head, the approximate scale over which 
equipartition should hold \citep{kc84}.  
This assumes the flow is subsonic, being confined  on the trailing side by
the high SNR pressure.  This assumption seems reasonable, as in the absence of
such confinement, it is hard to understand why this pulsar, which has space
velocity not much larger than the average pulsar \citep[e.g.][]{ll94}, should have
such a strking ram-pressure confined PWN.
Hence, the X-ray tail behind \psr\
is unlikely to be a trail in the conventional sense, in that it cannot be
synchrotron emission from pulsar wind particles just  left behind after the
passage of the pulsar.  

Rather, assuming the spectrum of the tail is indeed non-thermal,
freshly shocked wind particles must be 
continuously fed eastward with a velocity much larger than the 
pulsar space velocity, $v_f \gg v_t$.  
This is similar to the picture suggested for a putative X-ray tail
behind PSR~B1929+10 \citep[see][although that tail has not been
confirmed]{wlb93} and a cometary-shaped PWN in the
LMC SNR N157B \citep{wg98b}.  However, Lyutikov (in preparation) has 
developed an alternative model of the structure of the
tail region.
In this model, the pulsar wind is shocked near the pulsar
and forms a subsonically expanding tail confined by ISM ram pressure
(without forming a de Laval nozzle
as suggested
by \citeauthor{wlb93}~\citeyear{wlb93}). The typical velocities
 of the flow are weakly relativistic near the
pulsar, decreasing downstream as the tail expands.
At its end, the tail flow finally forms a  pressure confined expanding
bubble,
presumably G5.27$-$0.90.

We can constrain the flow velocity $v_f$ of the
wind particles in the tail by noting that it must be high
enough to continuously supply particles given their cooling
times.  Thus, the flow time $t_f \lapp t_s$.  Assuming  that the
magnetic field reaches its equipartition value near the bow-shock
head and that this value holds for the approximately
 one-dimensional  tail region too, we
find $t_s \gapp 70 (E / 1 \; {\rm keV})^{-1/2} (B_{\rm eq} / 70 \; \mu{\rm G})^{-3/2}$~yr.  As the tail
extends
to 0.48~pc for $d=5$~kpc, this implies 
$v_f \gapp 6800 (d/5 \; {\rm kpc})(E / 1 \; {\rm keV})^{1/2} (B_{\rm eq} / 70 \; \mu{\rm
G})^{3/2}$~km~s$^{-1}$.

The tail emission has low flux.  The efficiency with
which the pulsar's $\dot{E}$ is converted into tail X-rays in the
2--8~keV band is only 0.00011$(d/5 \; {\rm kpc})^2$,
roughly half of the point-source efficiency.  This is in
contrast to other rotation-powered pulsars, like the Crab, whose X-ray nebular emission
is much brighter than the point-source output.  
Without timing information, we cannot rule out an ultra-compact
nebula as the source of the point-source emission.  But even so,
the total efficiency for conversion of spin-down energy to nebula
energy would be significantly less than that of most rotation-powered
pulsars \citep{bt97}. 

Several effects  may reduce the X-ray efficiency of
ram pressure confined PWNs.
First,  the efficiency of conversion of  $\dot{E}$ into
X-ray emitting particles may be lower since the reverse shock 
in the ram-pressure-confined PWNs is strong  only 
in the forward part of the head,
 subtending  a much smaller solid angle than in a
static PWN.
Second, the low X-ray efficiency is  expected if
the flow time of the relativistic plasma  through the tail
is shorter than  the  synchrotron life time. In this case
the particles will be able to emit only a small fraction of the  energy
that they acquired during acceleration at the reverse shock;
most of the energy of the wind will be spent on doing
work inflating the bubble at the end of the tail.
Indeed, a similar argument was put forth \citet{che00}
to explain the low efficiencies
of the Vela and CTB~80 pulsars, both of which exhibit bow-shock
morphologies and have relatively
flat spectra.  Finally, 
it is also possible that low surface brightness emission
from beyond the eastern tip of the observable X-ray tail, or even
from G5.27$-$0.90, have gone undetected in our
observation.  For example, emission from the direction of that nebula, having
X-ray surface brightness half of the ACIS-S3 background, would contribute
roughly two orders of magnitude more flux.  

We can compare the observed X-ray tail flux with that predicted from
the radio flux density and spectrum.  \citet{fk91} found that the
radio tail has a flat spectrum ($\alpha \simeq 0$, where $\alpha$ is
the energy spectral index), consistent with what we find at
X-ray energies (\S \ref{sec:obs}).  However, it is impossible to
produce a continuous spectrum for the tail between
the radio, where $f_{20~\mu{\rm eV}} = 2.4 \times
10^{-8}$~erg~cm$^{-2}$~s$^{-1}$~keV$^{-1}$ \citep{fk91},
and X-ray, where $f_{\rm 5\, keV} \approx 1.6 \times
10^{-14}$~erg~cm$^{-2}$~s$^{-1}$~keV$^{-1}$ (\S\ref{sec:obs})
bands.  Thus there
must be an intrinsic spectral break, which, even if it lies as low as in the
radio band (e.g. 5~GHz), demands an X-ray photon index $>1.8$, much
larger
than expected for a synchrotron cooling break, and
only marginally consistent with the data (\S\ref{sec:obs}).
Therefore, the broadband spectrum has at
least two inflection points between the radio and X-ray bands, i.e.
the emission in the two bands originates from separate populations
of accelerated particles.  The same problem arises in the Crab nebula
\citep[e.g. see][]{kc84}.

\section{The Supernova Remnant}

We have detected no X-ray emission from SNR~G5.4--1.2.  Assuming
an association,
if the pulsar's characteristic age is a good estimate of the true age,
then the SNR is young ($t_{\rm SNR} \sim 16$~kyr), and
we might expect thermal X-ray emission from hot gas behind
the SNR shock.  
We considered a $4' \times 4'$ region of the SNR
centered on coordinates (J2000) RA 18$^{\rm h}$~01$^{\rm m}$~18$^{\rm s}$, 
DEC --24$^{\circ}$~51$'$~10$''$.
A background correction was applied using a region having the same
dimensions next to the source region, scaling for the different
effective areas.  We find no excess emission in
this region, with a 5$\sigma$ upper limit on the count rate of
0.017~cts~s$^{-1}$ in the range 0.3--8.0~keV.  
Assuming $d=5$~kpc, the shock velocity in the
Sedov solution should be $\approx 600$~\kms, corresponding to a shock
temperature $kT \approx 0.5$~keV.  Using a Raymond-Smith model with
$kT = 0.5$~keV and $N_H = 3\times10^{22}$~cm$^{-2}$, and assuming
the area under consideration encloses a volume
$\sim6\times10^{57}$~cm$^3$, the limit on the count rate in this region
corresponds to $n_e n_H < 0.15$~cm$^{-6}$, where
$n_e$ and $n_H$ are the electron and hydrogen densities, respectively.
Assuming a composition of pure ionized hydrogen, the upper limit on the
pre-shock density is $n_H < 0.1$~cm$^{-3}$.  This result is not a
strong function of the assumed temperature:  for $kT = 0.2$(5)~keV,
we find $n_H < 0.50 (0.04)$~cm$^{-3}$. These
limits are unconstraining.  If the system has an age of
16~kyr, the inferred density is $n_H \sim 0.003$~cm$^{-3}$
\citep{fk91}.
If the SNR is much older than indicated by the pulsar's spin-down
\citep{gf00}, then the SNR
expansion speed is
$<100$~\kms, and any X-ray emission from the SNR is expected to be 
too faint to detect for this distance and absorption.

We thank M. Roberts for useful discussions.
This work is supported by SAO grant GO0-1133A, and by
NASA LTSA grant NAG5-8063 and NSERC grant Rgpin 228738-00
 to V.M.K. E.V.G is supported by the NASA LTSA grant NAG5-22250.
B.M.G. acknowledges the support of a Hubble Fellowship
awarded by STScI.

\clearpage
\begin{deluxetable}{lcc}
\tablenum{1}
\tablewidth{310pt}
\tablecaption{Spectral Parameters for \psr\ for Two Trial Models\tablenotemark{a}.\label{ta:spectrum} }
\footnotesize
\tablehead{
\colhead{} & \colhead{Power Law} & \colhead{Thermal Bremsstrahlung}} 
\startdata
$N_H$ ( $10^{22}$~cm$^{-2}$) & 3.5$_{-1.1}^{+1.3}$  & 3.3$_{-1.0}^{+0.76}$ \\
$\Gamma$\tablenotemark{b} & 1.6$_{-0.5}^{+0.6}$  & ... \\
$kT$ (keV) & ... & 16$_{-8}^{+\infty}$ \\
Absorbed Flux\tablenotemark{c} & $5.4^{+0.4}_{-0.5}$ & $5.1^{+0.4}_{-0.5}$ \\
Unabsorbed Flux\tablenotemark{c} &
$6.9^{+0.5}_{-0.6}$  &    $6.6^{+0.5}_{-0.6}$     \\
$\chi^2_{\nu}$ (for 24 DoF) & 1.24 & 1.26 \\
\enddata
\tablenotetext{a}{Uncertainties reported are at the 90\% confidence level.}
\tablenotetext{b}{Photon index.}
\tablenotetext{c}{In $10^{-13}$ erg~cm$^{-2}$~s$^{-1}$ in the 2--10 keV band.}
\end{deluxetable}

\clearpage

\clearpage



\begin{thebibliography}{25}
\expandafter\ifx\csname natexlab\endcsname\relax\def\natexlab#1{#1}\fi


\bibitem[{Becker \& Tr\"{umper}(1997)}]{bt97}
Becker, W. \& Tr\"{umper}, J. 1997, A\&A, 326, 682

\bibitem[{Burke {et~al.}(1997)Burke, Gregory, Bautz, Prigozhin, Kissel,
  Kosicki, Loomis, \& Young}]{bgb+97}
Burke, B.~E., Gregory, J., Bautz, M.~W., Prigozhin, G.~Y., Kissel, S.~E.,
  Kosicki, B.~N., Loomis, A.~H., \& Young, D.~J. 1997, IEEE Transac. Elec.
  Devices, 44, 1633

\bibitem[{Caswell {et~al.}(1987)Caswell, Kesteven, Komesaroff, Haynes, Milne,
  Stewart, \& Wilson}]{ckk+87}
Caswell, J.~L., Kesteven, M.~J., Komesaroff, M.~M., Haynes, R.~F., Milne,
  D.~K., Stewart, R.~T., \& Wilson, S.~G. 1987, MNRAS, 225, 329

\bibitem[{{Chevalier}(2000)}]{che00}
{Chevalier}, R.~A. 2000, ApJ, 539, L45

\bibitem[{Cordes {et~al.}(1993)Cordes, Romani, \& Lundgren}]{crl93}
Cordes, J.~M., Romani, R.~W., \& Lundgren, S.~C. 1993, Nature, 362, 133

\bibitem[{Deutsch(1999)}]{deu99}
Deutsch, E.~W. 1999, Astron. J., 118, 1882


\bibitem[{Frail {et~al.}(1994)Frail, Kassim, \& Weiler}]{fkw94}
Frail, D.~A., Kassim, N.~E., \& Weiler, K.~W. 1994, Astron. J., 107, 1120

\bibitem[{Frail \& Kulkarni(1991)}]{fk91}
Frail, D.~A. \& Kulkarni, S.~R. 1991, Nature, 352, 785

\bibitem[{{Gaensler} \& {Frail}(2000)}]{gf00}
{Gaensler}, B.~M. \& {Frail}, D.~A. 2000, Nature, 406, 158


\bibitem[{Kennel \& Coroniti(1984)}]{kc84}
Kennel, C.~F. \& Coroniti, F.~V. 1984, ApJ, 283, 710

\bibitem[{Lyne \& Lorimer(1994)}]{ll94}
Lyne, A.~G. \& Lorimer, D.~R. 1994, Nature, 369, 137


\bibitem[{Manchester {et~al.}(1985)Manchester, D'Amico, \& Tuohy}]{mdt85}
Manchester, R.~N., D'Amico, N., \& Tuohy, I.~R. 1985, MNRAS, 212, 975

\bibitem[{Manchester {et~al.}(1991)Manchester, Kaspi, Johnston, Lyne, \&
  D'Amico}]{mkj+91}
Manchester, R.~N., Kaspi, V.~M., Johnston, S., Lyne, A.~G., \& D'Amico, N.
  1991, MNRAS, 253, 7P

\bibitem[{{Marsden} {et~al.}(2001){Marsden}, {Lingenfelter}, \&
  {Rothschild}}]{mlr01}
{Marsden}, D., {Lingenfelter}, R.~E., \& {Rothschild}, R.~E. 2001, \apjl, 547,
  L45

\bibitem[{Monet {et~al.}(1996)Monet, Bird, Canzian, Harris, Reid, Rhodes, Sell,
  Ables, Dahn, Guetter, Henden, Leggett, Levison, Luginbuhl, Martini, Monet,
  Pier, Riepe, Stone, Vrba, \& Walker}]{mbc+96b}
Monet, D., Bird, A., Canzian, B., Harris, H., Reid, N., Rhodes, A., Sell, S.,
  Ables, H., Dahn, C., Guetter, H., Henden, A., Leggett, S., Levison, H.,
  Luginbuhl, C., Martini, J., Monet, A., Pier, J., Riepe, B., Stone, R., Vrba,
  F., \& Walker, R. 1996, USNO-SA2.0, U.S. Naval Observatory, Washington DC



\bibitem[{Wang \& Gotthelf(1998)}]{wg98b}
Wang, Q.~D. \& Gotthelf, E.~V. 1998, ApJ, 494, 623

\bibitem[{Wang {et~al.}(1993)Wang, Li, \& Begelman}]{wlb93}
Wang, Q.~D., Li, Z.-Y., \& Begelman, M.~C. 1993, Nature, 364, 127

\end{thebibliography}
\end{document}